\documentclass[twocolumn]{aastex63}
\usepackage{hyperref}
\usepackage{amsmath,amssymb,mathrsfs}
\usepackage{soul}
\usepackage{lineno}

\def \psr {\mbox{1RXS\, J141256.0$+$792204}}
\def \xmm {\emph{XMM-Newton}}
\def \fermi {\emph{Fermi}}
\def \nic {\emph{NICER}}

\def \cha {\emph{Chandra}}

\def \pdot {\dot P}

\def\nudot {\dot \nu}
\def\nuddot {\ddot \nu}

\def\msun{{\rm M}_{\odot}}

\def\ltsima{$\; \buildrel < \over \sim \;$}
\def\lsim{\lower.5ex\hbox{\ltsima}}
\def\gtsima{$\; \buildrel > \over \sim \;$}
\def\gsim{\lower.5ex\hbox{\gtsima}}

\graphicspath{{fig/}}

\defcitealias{Davies81}{DP81} 

\received{2021 October 25}
\revised{2021 October 27}
\accepted{2021 October 28}
\submitjournal{ApJ}

\shorttitle{Calvera, a Galactic halo X-ray pulsar?}
\shortauthors{Mereghetti et al.}

\begin{document}

\title{\nic\ study of pulsed thermal X-rays from Calvera: a neutron star born in the Galactic halo?}

\correspondingauthor{Sandro Mereghetti}
\email{sandro.mereghetti@inaf.it}

\author[0000-0003-3259-7801]{S.~Mereghetti}
\affiliation{INAF -- Istituto di Astrofisica Spaziale e Fisica Cosmica, Via A. Corti 12, I-20133 Milano, Italy}
 
\author[0000-0001-6641-5450]{M.~Rigoselli}
\affiliation{INAF -- Istituto di Astrofisica Spaziale e Fisica Cosmica, Via A. Corti 12, I-20133 Milano, Italy}
 
\author[0000-0002-1768-618X]{R.~Taverna}
\affiliation{Dipartimento di Fisica e Astronomia, Università di Padova, via F. Marzolo 8, I-35131 Padova, Italy}

\author{L.~Baldeschi}
\affiliation{INAF -- Istituto di Astrofisica Spaziale e Fisica Cosmica, Via A. Corti 12, I-20133 Milano, Italy}
\affiliation{Dipartimento di Fisica, Università degli Studi di Milano, Via Celoria 16, I-20133 Milano, Italy}

\author[0000-0002-8368-0616]{S.~Crestan}
\affiliation{INAF -- Istituto di Astrofisica Spaziale e Fisica Cosmica, Via A. Corti 12, I-20133 Milano, Italy}
\affiliation{Università dell’Insubria, Dipartimento di Scienza e Alta Tecnologia, Via Valleggio 11, I-22100 Como, Italy}

\author[0000-0003-3977-8760]{R.~Turolla}
\affiliation{Dipartimento di Fisica e Astronomia, Università di Padova, via F. Marzolo 8, I-35131 Padova, Italy}
\affiliation{MSSL, University College London. Holmbury St. Mary, UK}

\author[0000-0001-5326-880X]{S.~Zane} 
\affiliation{MSSL, University College London. Holmbury St. Mary, UK}

\begin{abstract}
Calvera (\psr) is an isolated neutron star detected only through its thermal X-ray emission. Its location at high Galactic latitude ($b=+37^{\circ}$) is unusual if Calvera is a relatively young pulsar,  as suggested by its spin period (59 ms) and period derivative ($3.2\times10^{-15}$ Hz s$^{-1}$). 
Using the Neutron Star Interior Composition Explorer (\nic), we obtained a phase-connected timing solution spanning four years which allowed us to measure the second derivative of the frequency $\nuddot=-2.5 \times 10^{-23}$ Hz s$^{-2}$ and to reveal timing noise consistent with that of normal radio pulsars. A magnetized hydrogen atmosphere model, covering the entire star surface, provides a good description of the phase-resolved spectra and  energy-dependent pulsed fraction. However, we find that a temperature map more anisotropic than that produced by a dipole field is required, with a hotter zone concentrated towards the poles. By adding two small polar caps, we find that the surface effective temperature and that of the caps are $\sim$0.1 and $\sim$0.36 keV, respectively. The inferred distance is $\sim$3.3 kpc. We confirm the presence of an absorption line at 0.7 keV associated to the emission from the whole star surface, difficult to interpret as a cyclotron feature and more likely originating from atomic transitions. We  searched for pulsed $\gamma$-ray emission by folding seven years of \fermi-LAT data using the X-ray ephemeris, but no evidence for pulsations was found. Our results favour the hypothesis that Calvera is a normal rotation-powered pulsar, with the only peculiarity of being born at a large height above the Galactic disk.
\end{abstract}



\section{Introduction}
\label{sec:intro}

\psr\  is an enigmatic  X-ray pulsar with properties that do not fit easily with those of the known classes of isolated neutron stars (NSs). 
It was discovered in the ROSAT All Sky Survey as a soft  X-ray source  with high X-to-optical flux ratio, qualifying it as an isolated NS candidate \citep{rut08}.   Its spectral properties resemble those of the small class of thermally emitting  NSs known as X-ray Dim Isolated Neutron Stars (XDINSs, also called ``Magnificent Seven'', see, e.g., \citealt{tur09} for a review).    
 
The XDINSs have long spin periods (3--17 s),  very soft X-ray spectra (blackbody temperatures  $kT_{\rm BB}\sim0.05$--$0.1$ keV)   often exhibiting broad absorption spectral lines, and  X-ray  luminosities of $10^{31}$--$10^{32}$ erg s$^{-1}$.   
\psr\  has a slightly hotter thermal spectrum  ($kT_{\rm BB}\sim0.2$ keV)   and,  like the XDINSs, it does not show  any sign of non-thermal X/$\gamma$-ray components  \citep{hal13} nor radio emission \citep{hes07}. For these reasons, it was initially considered as a possible  new member of the ``Magnificent Seven'' class and nicknamed ``Calvera''. However, it was later discovered that Calvera has a spin period of 59 ms \citep{zan11} and is spinning down at a rate   $\pdot=3.2\times10^{-15}$ s s$^{-1}$  \citep{hal13}. These timing parameters give a characteristic age $\tau_c = 2.9\times10^5$~yr and  a dipole magnetic field at  the equator $B_{\rm d}=4.4\times10^{11}$ G, that  do not fit with those of  the  XDINSs, which  have  $\pdot$ of a few 10$^{-14}$ s s$^{-1}$,  $\tau_c \sim$1--4 Myr, and  $B_{\rm d}\sim10^{13}-10^{14}$ G.

The X-ray luminosity of Calvera is poorly constrained owing to its unknown distance:   $L_X\sim1.3\times10^{32} d_{\rm kpc}^2$ erg s$^{-1}$. 
Its sky position (Galactic coordinates $l = 118^{\circ}$, $b = +37^{\circ}$) implies a height of $600\times d_{\rm kpc}$ pc above the Galactic plane.  
A proper motion of $69\pm26$ mas yr$^{-1}$ in a direction nearly perpendicular and away from the Galactic  plane, was measured with \cha\ \citep{hal15}.
The corresponding projected velocity is $\sim290\times d_{\rm kpc}$ km s$^{-1}$.   
If Calvera has a velocity typical of the bulk of radio pulsars and it was born within the disk scale-height of young massive stars, its distance should be smaller than a few hundreds parsecs. 
On the other hand, if Calvera is at a much larger distance, considering $\tau_c$ as an upper limit on its true age,  it must have been born at a height $\gtrsim 90 \times d_{\rm kpc}$ pc above the Galactic disk, possibly from the explosion of a high velocity runaway  star. 

Given its  large rotational energy loss rate of $6.1\times10^{35}$ erg s$^{-1}$, some non-thermal  X- or $\gamma$-ray emission is expected, but none is seen, with a limit  of $L_{\gamma}<8\times10^{31} d^2_{\rm kpc}$ erg s$^{-1}$  in the 0.1--300 GeV range \citep{hal13}.
While the absence of radio emission  could be accounted for by an unfavorable orientation, it is unlikely that the same applies to the $\gamma$-rays, since they are normally emitted by pulsars in very wide beams. If the distance is smaller than one kpc, the  $\gamma$-ray luminosity of Calvera is at least two orders of magnitude below that of pulsars with similar spin-down power.

Several authors also discussed  possible connections of Calvera with  the  class of X-ray sources known as central compact objects (CCOs).  CCOs are a small group of steady thermal X-ray sources, undetected at radio or $\gamma$-ray energies, and located at the center of supernova remnants \cite[see, e.g.,][]{del17}.  Spin periods in the range 0.1--0.4 s and very small $\pdot$ values have been measured in three of them,  implying the presence of neutron stars with dipole fields of only (0.3--1)$\times10^{11}$ G. 
The lack of an associated SNR  led to consider Calvera as a possible ``orphaned''   \citep{zan11,hal11} or aged \citep{hal13}  CCO. The low magnetic field inferred for some CCOs might result from burial of an initially stronger field caused by the fall-back of part of the supernova ejecta \citep{ho11,vp12,tf+16}. As the CCO evolves, the magnetic field would re-emerge on a timescale of $\sim$10$^4$ yr while the SNR fades, giving rise to objects with properties similar to those of Calvera.

Here we report on recent X-ray observations of Calvera obtained with the Neutron Star Interior Composition Explorer (\nic ) instrument. We computed a neutron star model atmosphere and applied it to derive constraints on the Calvera geometry and temperature distribution through an analysis of phase-resolved spectra and pulse profiles. 
Using the \nic\ X-ray data, we also derived a new phase-connected timing solution spanning almost four years and used this ephemeris to search for pulsed $\gamma$-ray emission in the   \fermi-LAT data. 

\section{Observations and data analysis} 
\label{sec:data analysis}

We analyzed all the data of \psr\ available in the public archive of   \nic\ observations.  \nic\ is an instrument optimized for spectral and timing studies of neutron stars in the 0.2--12 keV range installed on the International Space Station  \citep{gen16}. It is based on 56 coaligned concentrator optics providing a total effective area of 1900 cm$^2$ at 1.5 keV. Each concentrator is coupled to a focal plane module (FPM) using a silicon drift detector. Four FPMs have not been working since the beginning of the mission.  \nic\ is a collimated instrument (i.e. it does not provide images) with a field of view 30 arcmin$^2$. The timing resolution is better than 300 ns.

The \nic\ data are split into individual observations (ObsId) spanning at most one day each. Due to the orbital and visibility constraints, each ObsId contains  a variable number of disjoint time intervals.
The observations of Calvera used here  were obtained  from 15/9/2017 to 26/2/2021 (see Table~\ref{tab:logtime}).
Results of the data obtained before October 2018 have been published by \citet{bog19}.  

We reduced the data with the \texttt{nicerdas} software (version 8c) including all  the most recently released patches and calibration files (\texttt{CALDB XTI20210707}).  
As a first step, we filtered the data using the program  \texttt{nicerl2} and the standard cuts. 
To exclude time intervals of high particle background, we selected data with  $K_p < 5$ and \texttt{COR\_SAX} $> 1.914~K_p^{0.684}+0.25$, where $K_p$ is an indicator of the effect of the Solar wind activity on the Earth magnetosphere and the condition on \texttt{COR\_SAX} excludes parts of the orbit in regions with low cut-off rigidity. To reduce the effect of optical loading in the detectors we also applied the filter  \texttt{FPM$\_$underonly$\_$count} $<200$. 

These filtered data (about 964 ks of  exposure) were used for the timing analysis. We barycentered the arrival times using the JPL DE430 solar system ephemeris and the source position computed for each observation taking into account the proper motion given in \citet{hal15}. 

To evaluate the background for the spectral analysis we used the 3C50 model \citep{rem21}. We first extracted the source and background spectra for all the individual observations, excluding time intervals with $|S0_{net}|>0.5$ counts s$^{-1}$ and $|hbg_{net}|>0.05$ counts s$^{-1}$, where $S0_{net}$  and  $hbg_{net}$  are the background-subtracted count rates in the 0.2--0.3 keV and 13--15 keV energy ranges, respectively.
By examining the resulting spectra at energies above 4 keV (where the source contribution is negligible), we found that in many cases the model overestimated the background, because it predicted a count rate much higher than the observed one. Therefore, we removed all the observations for which the predicted background count rate in the 4--12 keV range differed by more than 6$\sigma$ from the observed count rate in the same energy range. 
After all these data selections, resulting in a total exposure of 371 ks, we extracted the total spectrum of the source, as well as the corresponding background spectrum and response files. Moreover, in the total spectrum we removed the data from two  particularly noisy FPM (n. 34 and 43).
  
All the spectra were rebinned with a minimum of three channels per bin,  in order to avoid oversampling the instrument energy resolution,  and requiring  a signal significance of at least 5$\sigma$ in each bin. The spectral analysis was performed using the XSPEC software 12.12.0 including a systematic error of 2\%.  For the interstellar absorption we used the \texttt{tbabs} model.  Errors in the spectral parameters are given at the 90\% confidence level.

\subsection{Timing analysis}

\begin{table}
\caption{Timing parameters 
\label{tab:timing}}
 \begin{center}
\begin{tabular}{lcc}
\hline
\hline
 Parameter  &  $n=3$ & $n=4$     \\
\hline
T$_0$ (TDB) & \multicolumn{2}{c}{58$\,$260.83109832} \\
MJD range & \multicolumn{2}{c}{58$\,$014 -- 59$\,$272} \\
$\nu$ (Hz)      &  16.8921479996(2) & 16.8921479986(2)  \\
$\nudot$  ($10^{-13}$ Hz s$^{-1}$)   & --9.3948(1)  & --9.3965(2)                \\
$\nuddot$   ($10^{-23}$ Hz s$^{-2}$)   & --2.54(4)  &  --0.66(20)                  \\
$\dddot \nu$  ($10^{-31}$ Hz s$^{-3}$) & $\dots$ &  --5.9(6) \\
$\chi^2$/ dof    &   268.2 / 122 & 178.7 / 121           \\

\hline
\end{tabular}
\end{center}
\end{table}

\begin{figure}
  \includegraphics[trim=1cm 0cm 1cm 0cm,clip,width=1.\columnwidth]{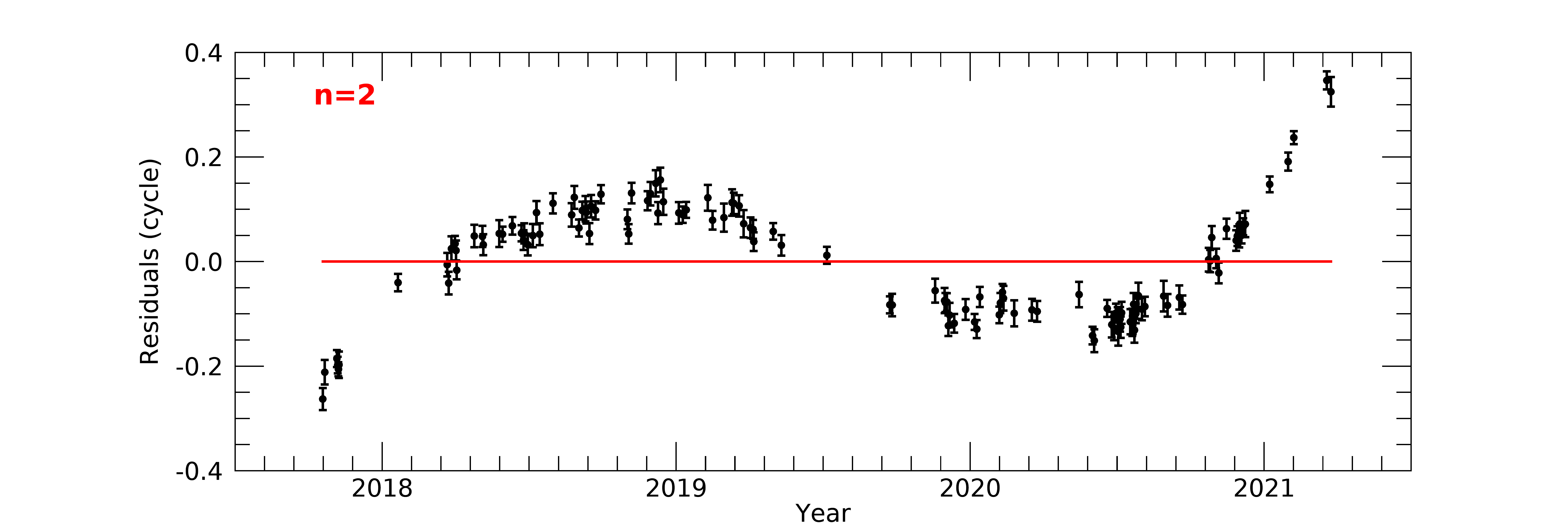}\\
  \includegraphics[trim=1cm 0cm 1cm 0cm,clip,width=1.\columnwidth]{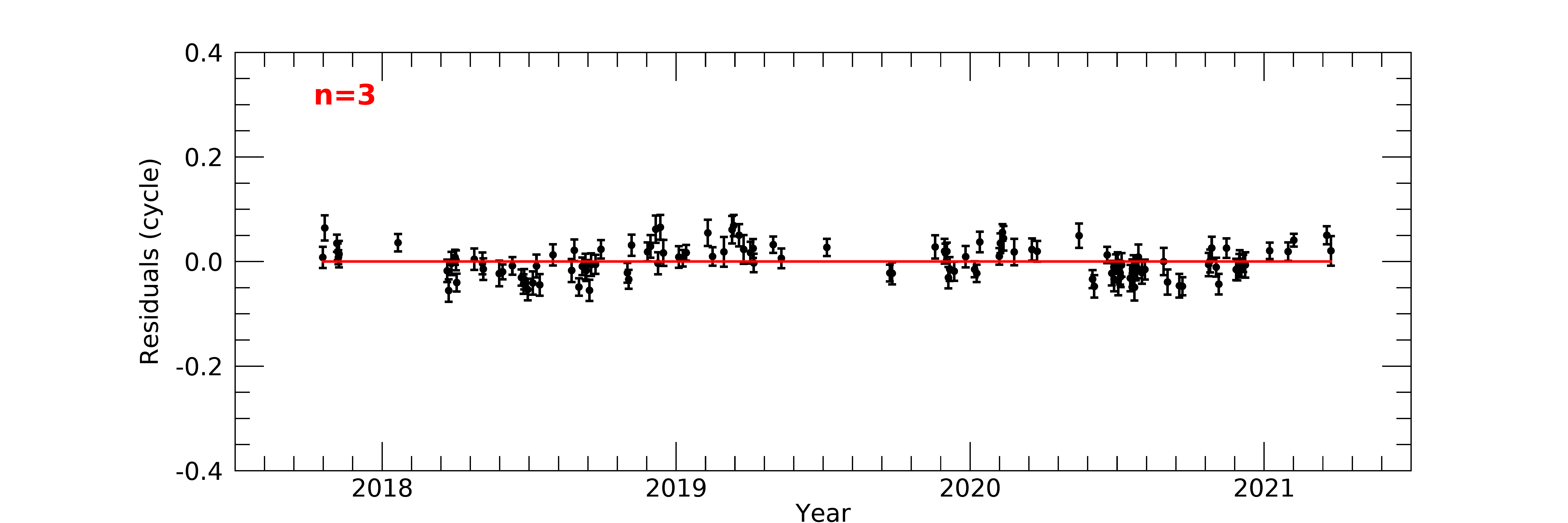}
  \includegraphics[trim=1cm 0cm 1cm 0cm,clip,width=1.\columnwidth]{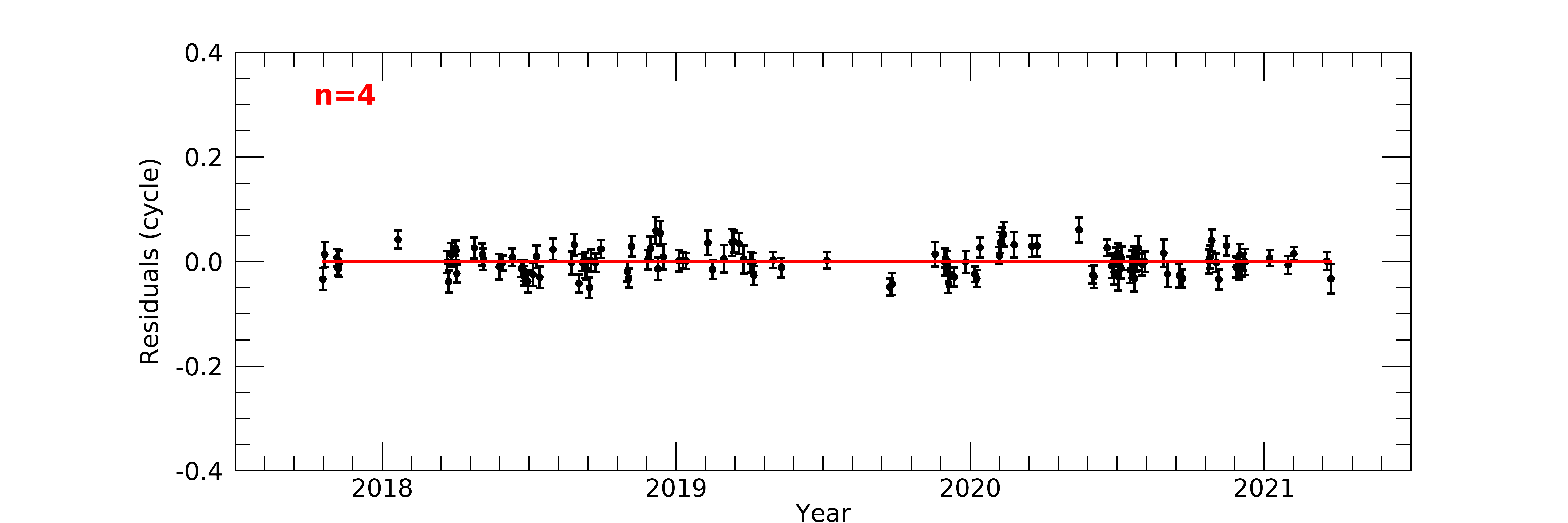}

  \caption{Residuals of the phase-connected timing solutions with only the $\nu$ and $\nudot$ terms (top) and adding the  $\nuddot$ term (middle). A further term, $\dddot\nu$, is included in the fit of the bottom panel.
 \label{fig:phasecon}
 }
\end{figure}

The X-ray pulsations of Calvera are clearly visible in most of the ObsId with net exposure longer than $\sim$2 ks. To accurately measure the source timing parameters we carried out a phase-coherent timing analysis. Briefly, this consists in fitting the pulse phases with the series expansion 

\begin{equation}
\phi(t) = \phi_0 + \nu (t-t_0) + \frac{1}{2} \dot\nu (t-t_0)^2 + \frac{1}{6} \ddot\nu (t-t_0)^3 + \dots 
\label{eq:phase}
\end{equation}
 \noindent
where $\nu$, $\nudot$, $\nuddot$, \ldots\ are the spin frequency and its time derivatives. The pulse phases were derived by fitting a sine function to the folded light curves in the 0.4--2 keV range. We started by fitting the closely spaced data of May 2018, and then gradually added more and more observations as the decreasing errors on the fit parameters allowed us to keep track of the number of pulse counts\footnote{ For each added time interval, we checked that the 3$\sigma$ uncertainty of the phase extrapolated from the previous iteration was smaller than 0.5.}. We grouped the ObsId in order to have at least 5 ks in each temporal bin, but we restricted the time span of each group to less than seven days. The resulting ObsId grouping is given in the last column of Table~\ref{tab:logtime}.
We started the phase fitting with only the $\nu$ and $\nudot$ terms, obtaining values consistent with those of \citet{bog19}. However, when we added the data of the second half of 2019, the fit residuals indicated the need of the second and then of the third derivative of the frequency. Figure~\ref{fig:phasecon} shows the residuals of the phase-connected timing solutions using the terms of degree $n=2$, $3$ and $4$ in Equation~(\ref{eq:phase}). The best-fit parameters, for the cases of the cubic and of the quartic solutions, are given in Table~\ref{tab:timing}.

\begin{figure}
  \includegraphics[trim=0cm 0.5cm 0cm 0.5cm,clip,width=1.\columnwidth]{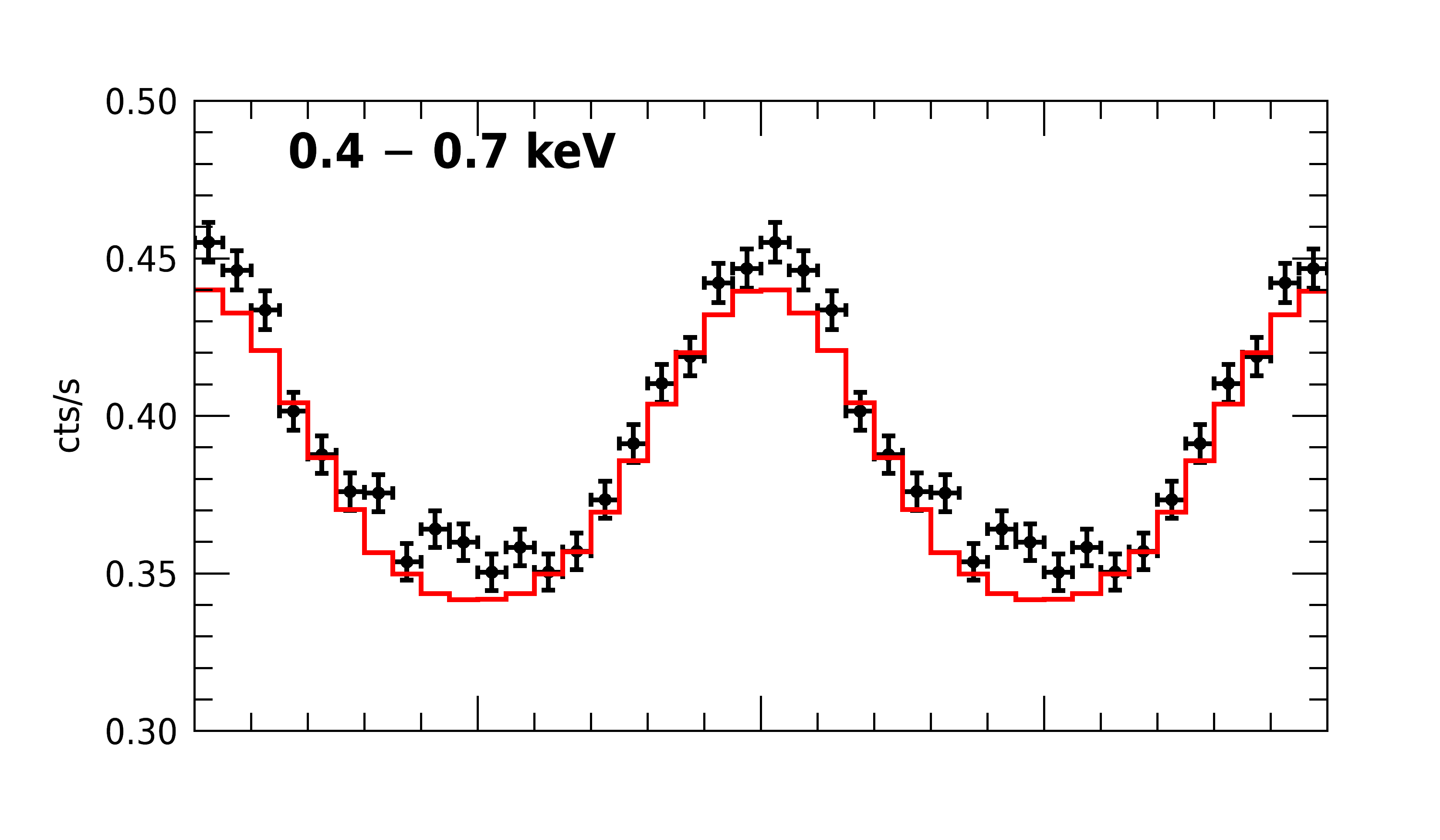}
  \includegraphics[trim=0cm 0.5cm 0cm 0.5cm,clip,width=1.\columnwidth]{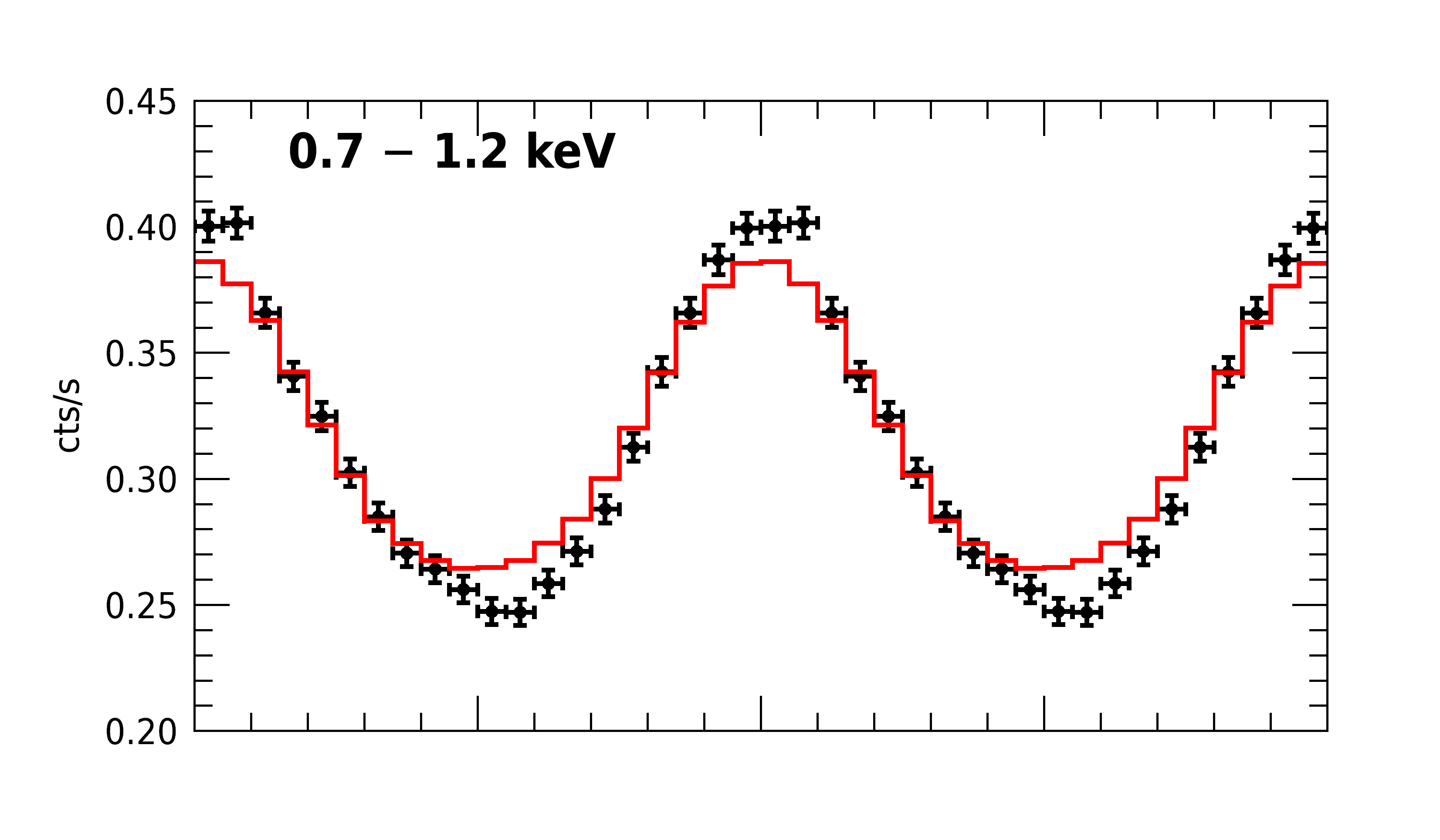}\\
  \includegraphics[trim=0cm 0.5cm 0cm 0.5cm,clip,width=1.\columnwidth]{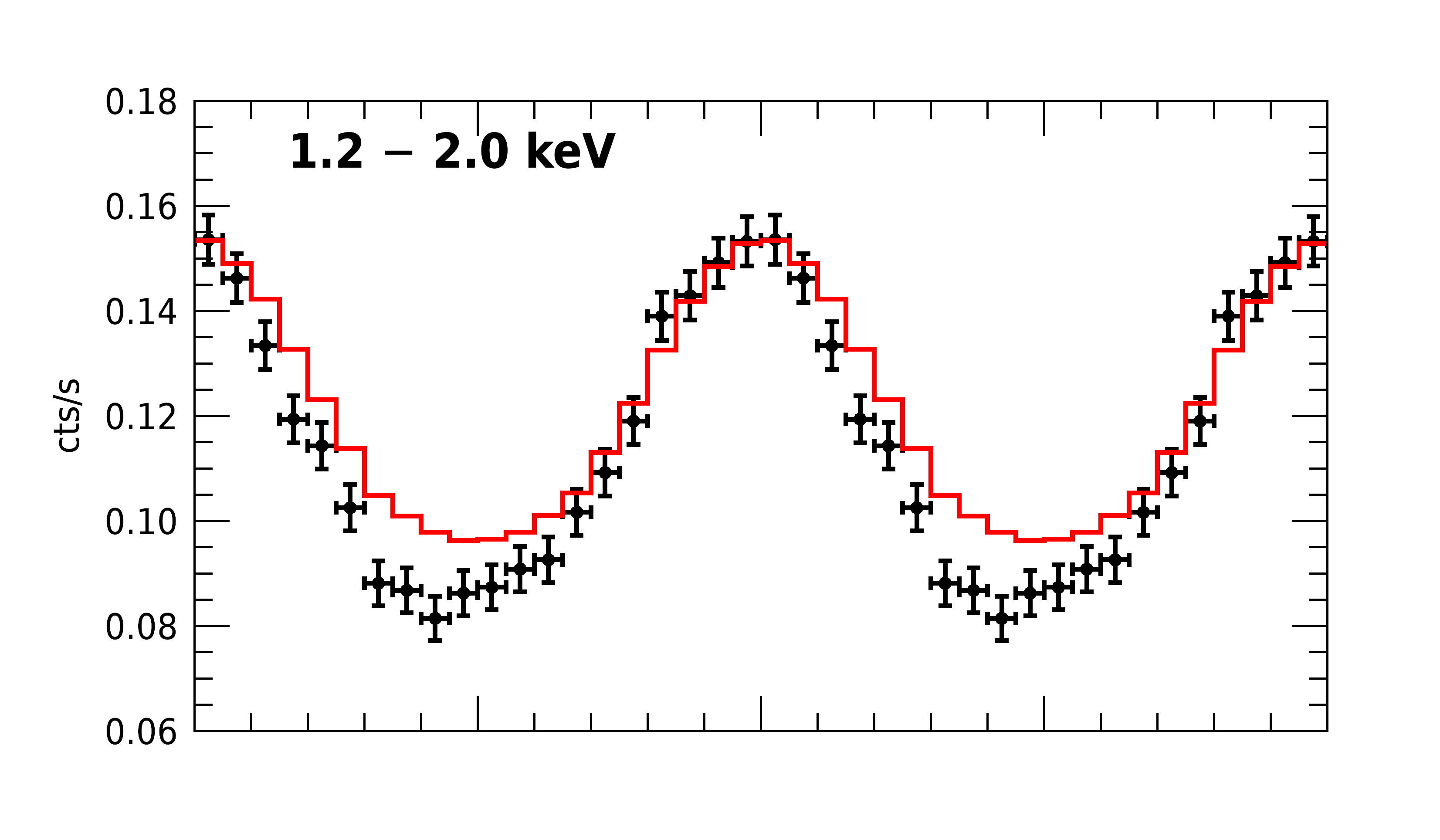}
  \includegraphics[trim=0cm 0cm 0cm 0.5cm,clip,width=1.\columnwidth]{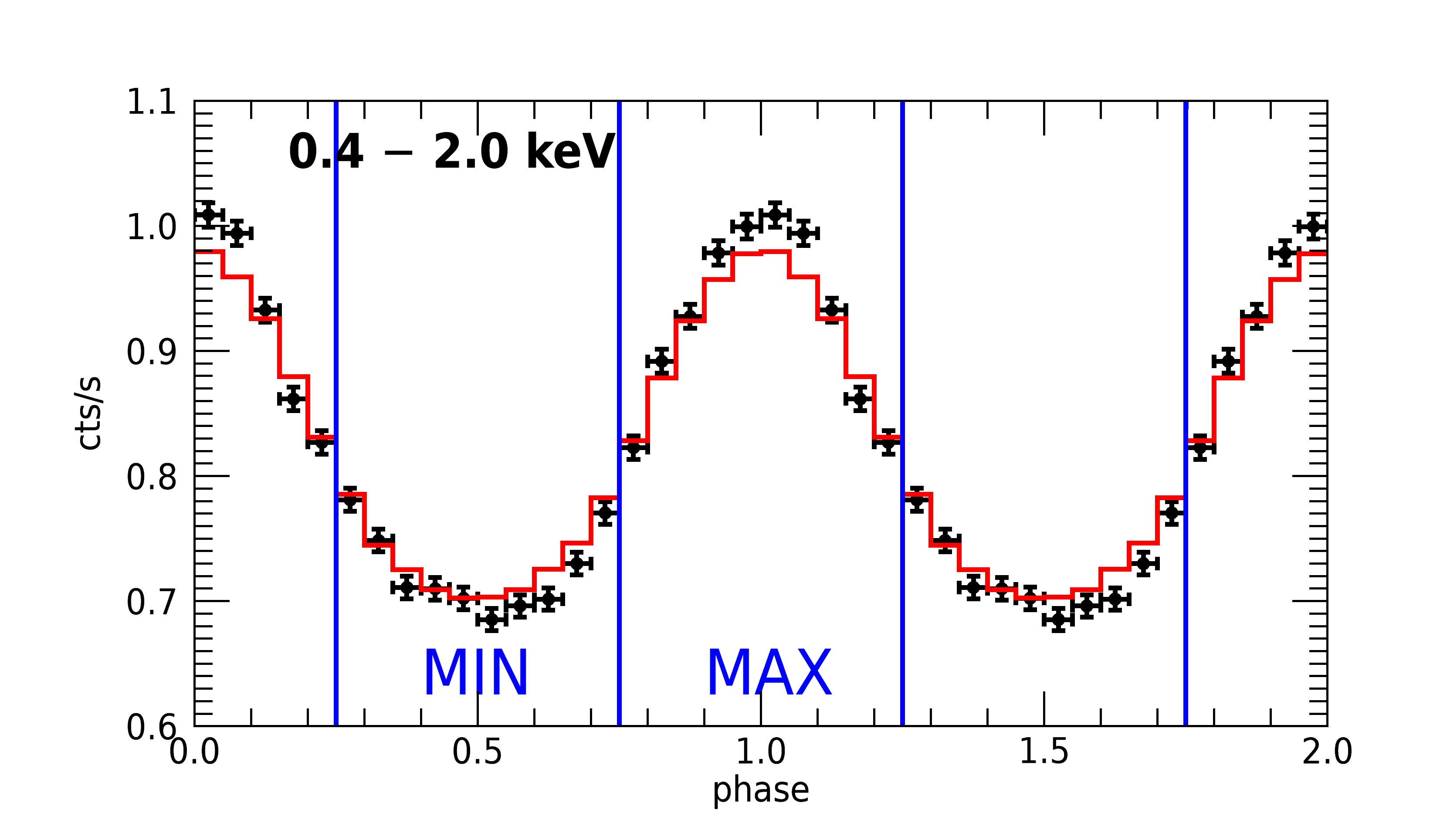}
  \caption{Background-subtracted pulse profiles in three energy ranges (0.4--0.7 keV, upper panel; 0.7--1.2 keV, middle panel; 1.2--2 keV, lower panel), and in the total energy range 0.4--2 keV (bottom panel). The red lines represent the pulse profiles computed with our best fit spectral model (see Section~\ref{sec:spectral}). The blue vertical lines show the phase intervals  used to extract the spectra of the maximum and of the minimum.
 \label{fig:lc}
 }
\end{figure}

Figure~\ref{fig:lc} shows the folded light curve in the 0.4--2 keV energy range (bottom panel) and in three representative energy ranges. The pulse profile is smooth and single-peaked, and the pulsed fraction (PF)\footnote{Defined as (max(CR)--min(CR))/(max(CR)+min(CR)), where CR is the background-subtracted count rate.} increases as a function of the energy: 0.13(1) between 0.4--0.7 keV, 0.24(1) between 0.7--1.2 keV, and 0.31(3) between 1.2--2 keV.

\subsection{Spectral analysis} \label{sec:spectral}

\begin{figure}
    \centering
    \includegraphics[width=0.5\textwidth]{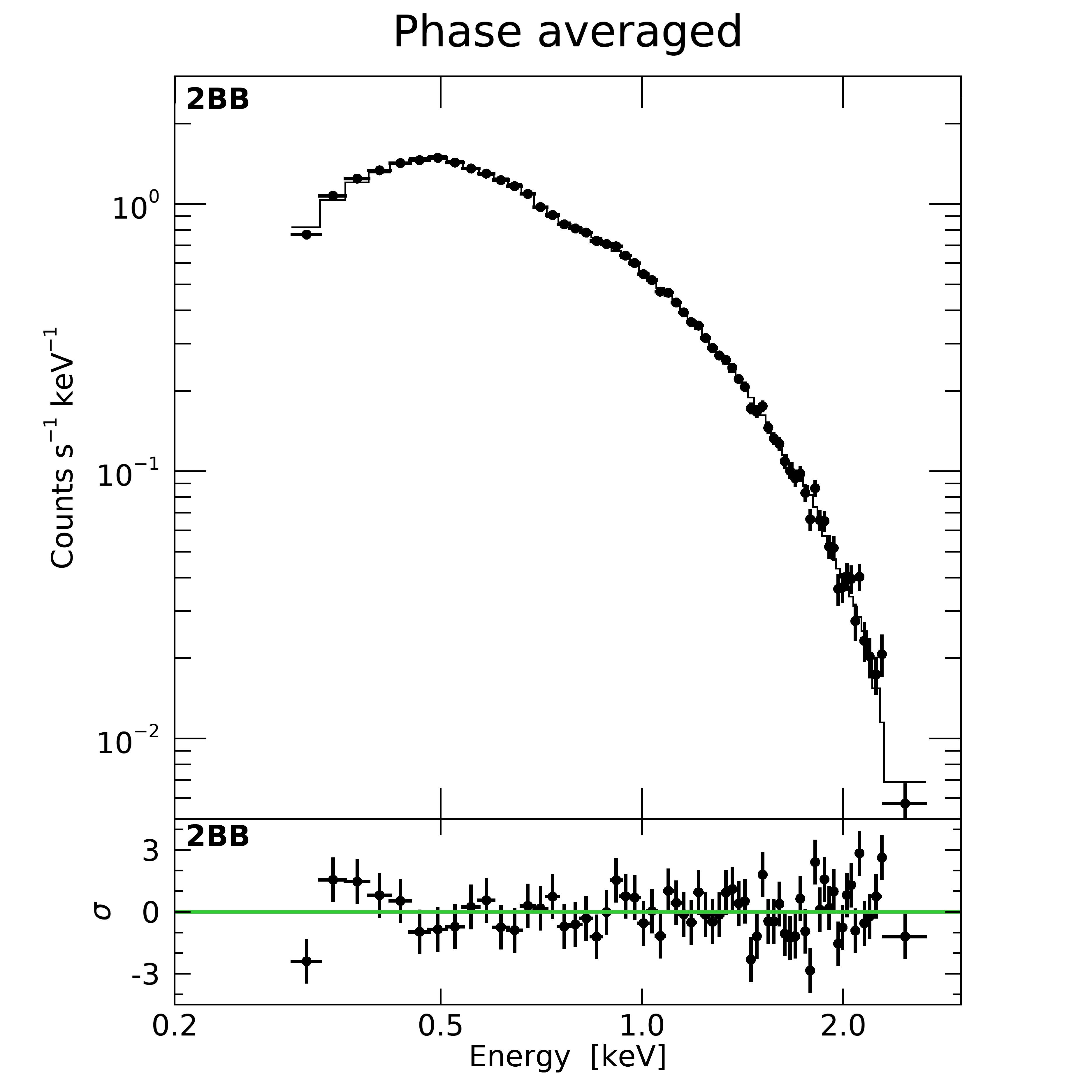}
    \caption{Spectrum of Calvera fitted with two blackbodies + absorption edge (parameters in Table \ref{tab:sp}).}
    \label{fig:sp}
\end{figure}

For the spectral analysis of Calvera we considered the energy range 0.3--2.5 keV.
We found that the phase-averaged spectrum can not be fitted by simple single-component models, such as a blackbody or a power law ($\chi^2 > 1900$ for 63 degrees of freedom, dof).
A better fit  was found with the sum of two blackbody models with temperatures $kT_1\sim0.15$ keV and  $kT_2\sim0.29$ keV, and an absorption edge at $E_{\rm edge}=0.68$ keV  ($\chi^2/$dof=85.3/59). The best fit is plotted in Figure~\ref{fig:sp} and all the  fit parameters are given in Table~\ref{tab:sp}.  
Models in which the edge is replaced by two lines were also acceptable, but they involve  a larger number of free parameters and will not be considered in the following.

\begin{table}
\caption{Results of phase-averaged spectroscopy
\label{tab:sp}
}
 \begin{center}
 \begin{tabular}{lc}
 \hline
 \hline
\smallskip

 Parameter         &     Value  \\
\hline
\smallskip

$N_{\rm H}$ ($10^{20}$ cm$^{-2}$)           & $1.7\pm0.4$        \\
\smallskip
$kT_{1}$     (keV)          &   $0.148\pm0.008$ \\
\smallskip
$R_{1}^{(a)}$  (km)        &    $1.24_{-0.11}^{+0.13}$   \\
\smallskip

$kT_{2}$     (keV)          &   $0.286_{-0.012}^{+0.016}$   \\
\smallskip
$R_{2}^{(a)}$  (km)        &    $0.26\pm0.04$     \\
\smallskip
$E_{\rm edge}$  (keV)     &   $0.68\pm0.01$  \\
\smallskip
$\tau$                    & $0.30\pm0.05$  \\
\smallskip
$F^{(b)}$  ($10^{-12}$ erg cm$^{-2}$ s$^{-1}$) &  $1.20\pm 0.09$    \\
\smallskip
$\chi^2$/ dof            &       85.3 / 59         \\

\hline
\end{tabular}
\end{center}
\textbf{Notes.} All the errors are at the 90\% c.l. for a single interesting parameter.

$^a$ Blackbody emission radius at infinity, for $d=1$ kpc.

$^b$ Observed flux  0.2--10 keV.

\end{table}

\begin{figure}
    \centering
    \includegraphics[width=0.5\textwidth]{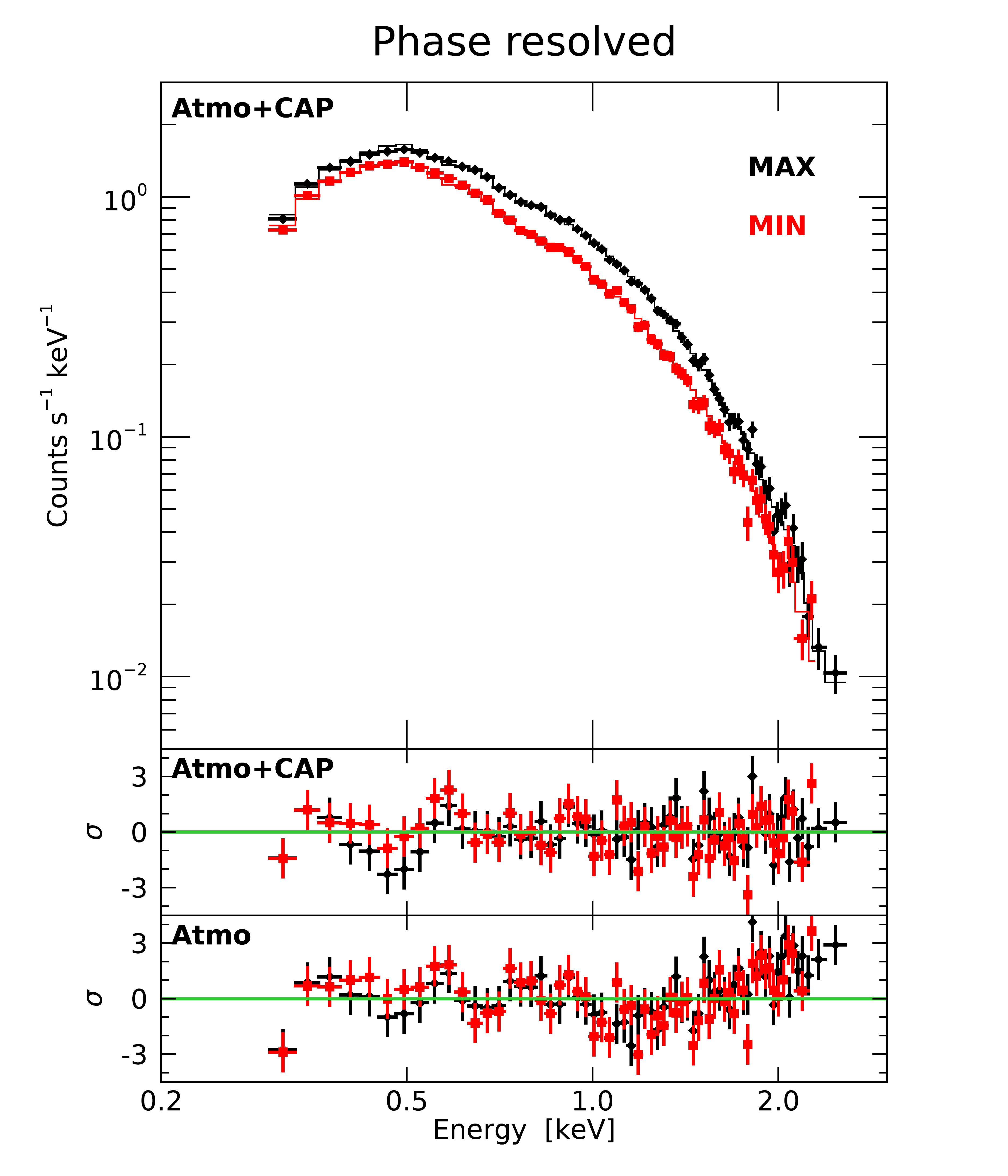}
    \caption{Spectra of the maximum (black) and minimum (red) fitted with our best model (parameters in Table \ref{tab:sp2}).}
    \label{fig:sp2}
\end{figure}

Given the high statistical quality of the \nic\ data, rather than trying other thermal models available in XSPEC on the phase-averaged spectrum, we proceeded directly to phase-resolved spectroscopy. For this, we used a neutron star atmosphere model  computed ad hoc for the Calvera case as follows.

In order to solve the radiative transfer in the plane-parallel NS atmospheric layer in local thermodynamic equilibrium, we used the code presented by \citet{llo03}, which includes (magnetic) bremsstrahlung and Compton scattering in the source term, with proper expressions for the opacities. The effects of strong magnetic fields are properly dealt with by considering photons polarized both in the ordinary mode (i.e. with the photon electric field oscillating 
in the $\boldsymbol{kB}$ plane, with $\boldsymbol{k}$ and $\boldsymbol{B}$ the photon propagation direction and the star magnetic field, respectively) and in the extraordinary mode (i.e. with the photon electric field oscillating in the $\boldsymbol{k}\times\boldsymbol{B}$ direction). Each run of the code requires as input parameters the surface gravity $g$ and temperature $T$, as well as the magnetic field strength and inclination with respect to the surface normal at the emission point, and returns the emerging intensities in the two polarization modes as functions of the photon energy and direction with respect to the local normal. The code is designed to treat only pure hydrogen, fully-ionized atmospheres. 

As for the mass and radius of the NS, we adopted the values of  $M=1.36~\msun$ and  $R=13$ km (consistent with the equation of state as in e.g., \citealt{2016PhR...621..127L} and references therein); this corresponds to a surface gravity $\log g =14.11$.
We computed the model atmosphere for ten surface patches, equally spaced in $\mu=\cos\theta$, where $\theta$ is the magnetic colatitude. The magnetic field is assumed to be a dipole with polar\footnote{$B_{\rm d}$ inferred from the timing measurements given above is the values at the star equator.} strength $B_{\rm p}=10^{12}$ G. The polar value of the temperature (as measured at infinity) has been varied by taking six equally spaced values in the range $0.08\, \mathrm{keV}\leq T_{\rm atm}\leq 0.16\, \mathrm{keV}$, and, once $T_{\rm atm}$ is fixed, the temperature distribution in the rest of the NS surface follows that of a dipole, with the corrections by \cite{pot15}.
Besides $B_{\rm p}$ (which is held fixed)  and $T_{\rm atm}$, each model is characterized by the two angles $\chi$ and $\xi$, which measure the inclination of the line-of-sight and of the dipole axis with respect to the rotation axis, respectively; they are sampled by means of a $19\times 19$ equally-spaced grid ranging, for both angles, from $0^{\circ}$ to $90^{\circ}$.

The spectrum at infinity was computed by collecting all the contributions from the patches that are in view at a certain rotational phase, accounting for general relativistic effects. By integrating over pre-selected phase intervals, we could then obtain the model spectra corresponding to the maximum and minimum of the pulse profile (see the vertical blue lines in Figure~\ref{fig:lc}).  These models were used to simultaneously fit the \nic\ spectra and to derive a single set of the best-fit parameters $T_{\rm atm}$, $\chi$, $\xi$, and the model normalization. Given that the spectral models are computed assuming that the radiation comes from the whole surface, the normalization constant can then be univocally related to the star distance.

\begin{table}
\caption{Results of phase-resolved spectroscopy
\label{tab:sp2}
}
 \begin{center}
 \begin{tabular}{lcc}
 \hline
 \hline

\smallskip
 Parameter         &     Atmosphere plus &  Atmosphere plus \\
            &    H polar caps &  BB polar caps \\

\hline
\smallskip

$N_{\rm H}$ ($10^{20}$ cm$^{-2}$)           & $3.1\pm0.3$    &     $2.8_{-0.1}^{+0.3}$  \\
\smallskip
$kT_{\rm atm}$     (keV)          &   $0.103 \pm 0.002$   &  $0.107 \pm 0.003$ \\
\smallskip
$d^{(a)}$  (kpc)        &   $3.27_{-0.16}^{+0.17}$    & $3.59 \pm 0.22$ \\
\smallskip
$kT_{\rm PC}$     (keV)          &   $0.358_{-0.035}^{+0}$   &  $0.84_{-0.26}^{+0.16}$ \\
\smallskip
$R_{\rm PC}^{(b)}$  (m)        &     $340_{-28}^{+105}$  &   $63_{-16}^{+66}$\\
\smallskip
$\chi$  (deg)  &  $27_{-8}^{+13}$ & $25_{-5}^{+11}$ \\
\smallskip
$\xi$ (deg)  &    $70_{-10}^{+4}$ & $66 \pm 6$ \\
\smallskip
$E_{\rm edge}^{(c)}$  (keV)     &    $0.69_{-0.01}^{+0.01}$ / $0.69_{-0.01}^{+0.01}$ &  $0.69_{-0.01}^{+0.01}$ / $0.69_{-0.01}^{+0.01}$ \\
\smallskip
$\tau^{(c)}$                    &  $0.25_{-0.04}^{+0.04}$ / $0.30_{-0.04}^{+0.04}$ & $0.28_{-0.04}^{+0.04}$ / $0.32_{-0.04}^{+0.04}$ \\
\smallskip
$\chi^2$/ dof            &       148.99 / 117    & 150.19 / 117    \\
\hline
\end{tabular}
\end{center}
\textbf{Notes.} All the errors are at the 90\% c.l. for each single parameter.

$^a$ Distance implied by the atmosphere model normalization (for $R_{\rm NS}$ = 13 km adopted  in the model computation).

$^b$ Radius of polar cap emission component, for $d$ equal to the best fit value.

$^c$ Values at pulse maximum/minimum.
\end{table}

We found that  this model alone (\texttt{ATMO}, in the following) could not properly fit the spectra, due to the presence of an excess at high energy, in addition to the absorption feature already seen in the phase-averaged spectrum (see the bottom panel of Figure~\ref{fig:sp2}). The latter was modeled again with an absorption edge, while the high energy excess required an additional hotter component. In order to account for this, we assumed that two, small and hot, polar caps are present, the emission of which was modeled either with a blackbody or with the same magnetized hydrogen atmosphere adopted for the rest of the surface. This further (thermal) component  was added to the fit of the phase-resolved spectra, linking the values of $\chi$ and $\xi$ to those of the  \texttt{ATMO} component; the cap temperature, $T_\mathrm{PC}$, and radius, $R_\mathrm{PC}$, are left free to vary (being the caps at the poles, the magnetic field has strength $B_\mathrm p$ and is normal to the cap surface). In this way we could obtain good fits (see Figure~\ref{fig:sp2}) with the parameters reported in Table~\ref{tab:sp2}. The temperature and normalization of the emission from the whole star do not depend much on the emission model used for the polar caps (in particular a distance  $\sim$3.1--3.8 kpc is obtained in both cases), while the cap temperature is higher by a factor $\simeq$2.3 in the blackbody case. The parameters of the absorption edge were left free to vary as a function of the phase, resulting in  marginal evidence for a change in the optical depth. 

\begin{figure}
  \includegraphics[width=1.\columnwidth]{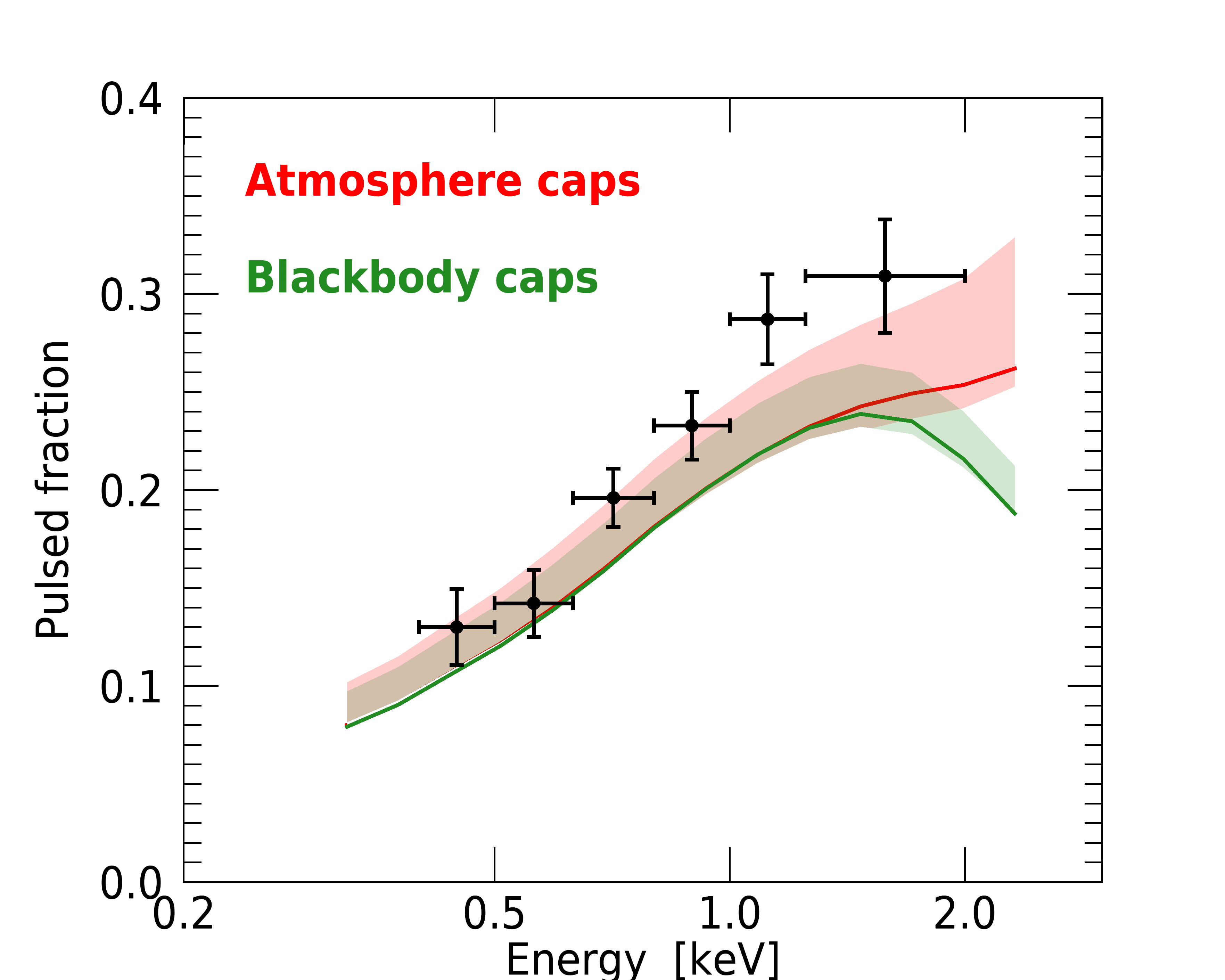}
  \caption{Pulsed fraction as a function of energy. The solid lines shows the pulsed fraction computed from the best fit spectral model (Table~\ref{tab:sp2}): an atmosphere model for the entire surface and the hot spots (red), or an atmosphere model for the entire surface and blackbody emission from the hot spots (green). The shadowed areas are obtained in both cases allowing the two angles $\xi$ and $\chi$ to vary within their $90\%$ confidence intervals. 
 \label{fig:pf}
 }
\end{figure}

Using the best-fit models obtained from the spectral analysis we can also compute the expected pulse profiles and the PF. Figure~\ref{fig:pf} shows the PF as a function of the energy in the case of two polar caps modeled with the magnetized hydrogen atmosphere (red solid line) and with the blackbody (green solid line). The two models predict the same PF below $\sim$1 keV, where the bulk of the emission 
comes from the (full-surface) \texttt{ATMO} component. The colored shadows have been computed letting $\xi$ and $\chi$ free to vary within their 90\% confidence level range as obtained from the spectral fit (Table~\ref{tab:sp2}). Within this range, we found that the better agreement with the observed values of the energy-dependent PF (black dots in Figure~\ref{fig:pf}) is obtained for $\xi\approx20^\circ$ and $\chi\approx60^\circ$. 
We then used these angles to compute the expected pulsed profiles in four energy ranges (red lines in Figure~\ref{fig:lc}).

\subsection{Timing analysis of \fermi-LAT data}

\begin{figure}
  \includegraphics[width=1.\columnwidth]{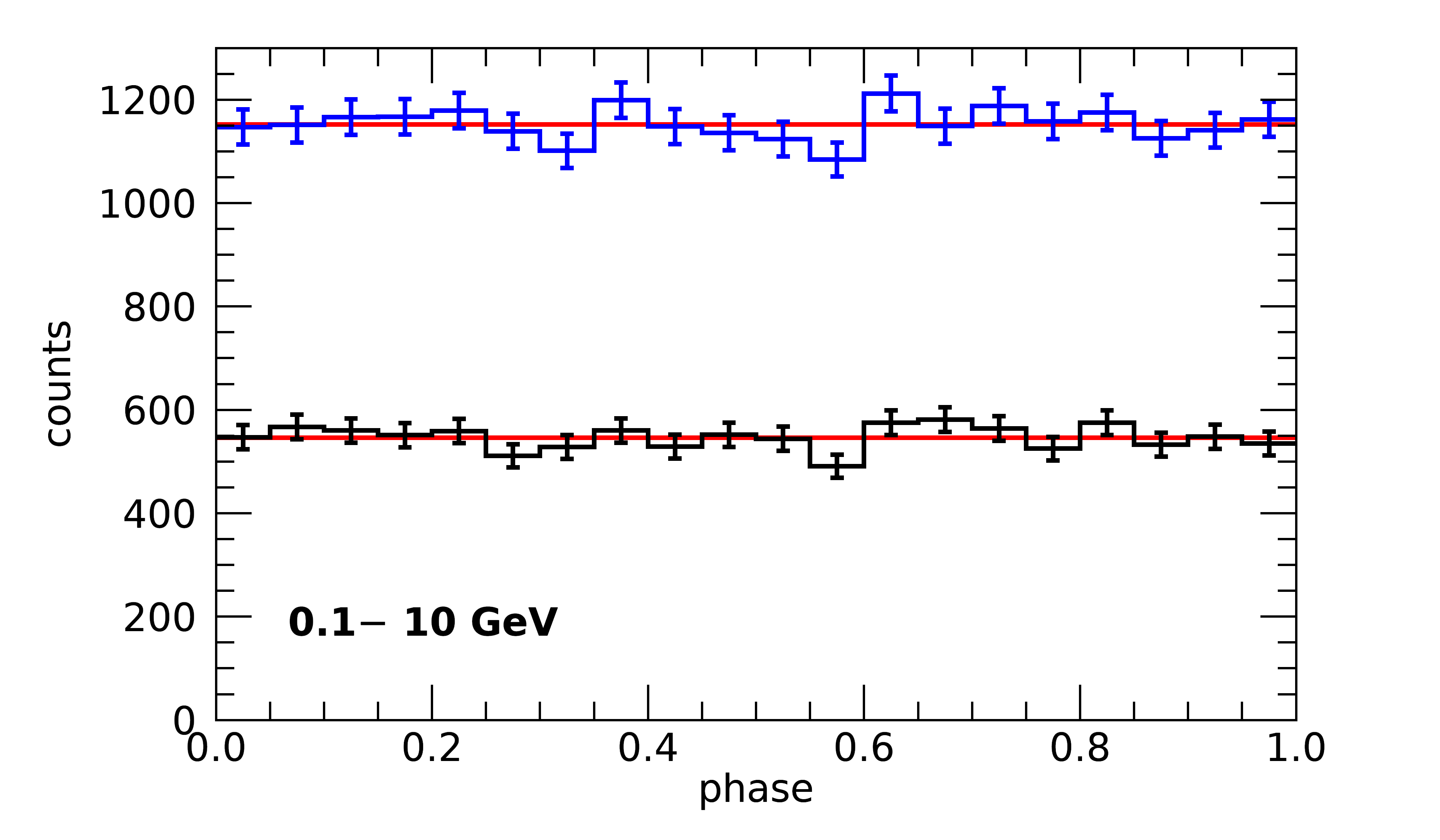}
  \caption{Folded light curves of \fermi-LAT data with energies in 0.1--10 GeV and times of arrival in the range 58$\,$014--59$\,$272 MJD (black dots) or 56$\,$658--59$\,$480 MJD (blue dots). Both are consistent with constant emission (red lines).
 \label{fig:lat}
 }
\end{figure}

Exploiting the phase-connected timing solution derived from the X-ray data it is possible to search for the presence of a pulsed signal in the \fermi-LAT \citep{Fermi} $\gamma$-ray data with a better sensitivity than in previous similar analysis.
We extracted Pass 8 \fermi-LAT events from a circular region with radius of $2^\circ$ centered at the position of Calvera and  barycentered their times of arrival taking into account the source proper motion, as done for the \nic\ data. 
We first considered only the events collected during the time period in which our timing solution is valid. Selecting events of the \texttt{SOURCE} class with energy  $E>0.1$ GeV and using the ephemeris of  Table~\ref{tab:timing}, we obtained the folded light curve shown in Figure~\ref{fig:lat} with black dots. This curve is consistent with a constant emission ($\chi^2=18.62$ for 19 dof).
We then expanded our time baseline to the whole period for which we could ensure the coherency of our timing solution, i.e. we included the events collected after January 2014. This resulted again in an unpulsed signal ($\chi^2=15.95$ for 19 dof, see Figure~\ref{fig:lat}, blue dots).
We also tried a different energy range ($E>0.3$ GeV) and/or class selection (\texttt{ULTRACLEANVETO}), but also in these cases no evidence for pulsations was found.

\section{Discussion} \label{sec:discussion}

Previous  studies of the Calvera X-ray  spectrum were based on  fits with (the combination of) single temperature thermal components, using either blackbody emission or some of the atmosphere models available in XSPEC. \citet{she09} obtained a good fit to the \cha\ phase averaged spectrum\footnote{The time resolution of the ACIS instrument used by these authors is 0.44 s, insufficient for phase-resolved spectroscopy.}   with a hydrogen atmosphere model (\texttt{NSA} in XSPEC) and either an absorption edge at 0.64 keV or a Gaussian emission line at 0.53 keV. They favored the latter interpretation on the basis of a higher $\chi^2$ improvement. Their best fit normalization, assuming emission from the whole star surface, implied a distance of 3.6 kpc, but no pulsations would be expected in this case. Single temperature thermal models could not fit the higher quality \xmm\ spectra \citep{zan11,hal13}.  The  model favored by \citet{zan11} consisted of the sum of two blackbodies, with temperatures of $\sim$0.15 and $\sim$0.25 keV, plus a possible absorption edge at $\sim$0.65 keV. If the colder thermal component is from the whole NS surface, the implied distance would be $\sim$5 kpc (for the two blackbody fit) or  $\sim$1.5--2 kpc (for the fit with two \texttt{NSA}). Contrary to the single-temperature fits, the models used by \citet{zan11} can  qualitatively explain the observed pulsations (the same is true for other publicly available atmospheric models explored by \citealt{shi16}), but  a quantitative modeling of the pulse profile was not carried out by these authors.
The phase-averaged spectrum from the \nic\ data obtained before October 2018 was fitted by   \citet{bog19} with either a blackbody plus power law or two blackbodies  with parameters similar to those of \citet{zan11}.  In both cases the inclusion of an absorption line at $\sim$0.77 keV and an emission line at $\sim$0.55 keV was needed. 

Our results are in general agreement with all the findings described above, but our model, having an intrinsically non-uniform temperature distribution,  has the advantage of naturally accounting for the observed pulsations. By simultaneously fitting the spectra of two different phase intervals (maximum and minimum), we could obtain some constraints on the angles $\chi$ and $\xi$\footnote{Due to the model symmetry, the best fit values of the two angles are interchangeable. In fact both  spectra and light curves are not sensitive to an exchange of $\chi$ into $\xi$. However,  this degeneracy can be broken by performing X-ray polarimetric observations, specifically by measuring the phase-dependent polarization angle of the X-ray signal.}.  

We found that two hotter polar spots, superimposed to the non-isotropic temperature distribution 
produced by the dipole field, are required to fit the spectra and the pulse profiles. The nature of these spots is uncertain. They can be due to a more complex surface temperature map, like that produced by a strong crustal toroidal field, of which our modeling is just an oversimplified description. In this case, in fact, the insulating effect of the toroidal field allows heat to be conducted to the surface only close to the poles, producing a temperature distribution which is more anisotropic than that induced by a pure dipole \cite[][]{gep06}. Alternatively, the spots can be  also produced by some form of external heating related to returning magnetospheric currents, as in radiopulsars \cite[][see also \citealt{tsyg17}]{sturr71,arons79}. If this is the case, it is unlikely that emission from the caps come from an atmosphere in radiative and hydrostatic equilibrium, like that we used to describe the rest of the surface. The radiative properties of bombarded atmospheres are not completely investigated as yet, but there are indications that the spectrum can be described by a single blackbody \cite[][]{denis19}. For this reason we modeled the contribution from the caps either with an atmosphere or a blackbody. On the other hand, the fact that the atmospheric model for the caps provides a better agreement with the observed trend of the PF with energy may be taken as indicative that particle bombardment is not substantial and indeed the hot polar region is created by a non-dipolar magnetic field in the crust.

We confirm the presence of a broad absorption feature at $\sim$0.7 keV. The absorption edge we used is the phenomenological model which requires the smallest number of parameters. The line is seen at all rotational phases and it falls at an energy range where most of the flux in our model is contributed by the emission from the entire star surface. This suggests that the line is not related to the polar caps emission component. If interpreted  as a cyclotron line from protons on the star surface, a field of $\sim$10$^{14}$ G is required, much larger than that inferred for the spin-down dipolar field ($B_{\rm p}= 2 B_{\rm d} \simeq 9\times10^{11}$ G). This is in contrast with what observed in the XDINSs, where the spin-down measure agrees rather well with the value of $B$ as derived from the energy of the broad absorption lines \citep{tur09}. We also note that the option to invoke small-scale, highly magnetized loops as responsible for a proton cyclotron line can work in sources where the feature is strongly phase-dependent \cite[as in the case of SGR 0418+5729 or some of the XDINSs\footnote{These are narrow absorption lines not to be confused with the broad ones discussed above.},][]{tiengoetal13,borgetal17}, but appears quite untenable for Calvera in the light of our present results. An explanation 
in terms of an electron cyclotron line is also questionable since the required magnetic field strength in this case is $\sim$7$\times 10^{10}$ G, about one order of magnitude smaller than the surface dipole value. This can be circumvented if the line forms at a height $\sim$2.5$R$ above the surface, so that the magnetic field has decayed to the required value. On the other hand, even then it would be hard to explain how a sufficiently large electron density can be maintained in the inner magnetosphere.
A more viable interpretation is that the feature is produced by atomic transitions.  H or He transitions in a field $\sim$10$^{12}$ G have a maximum energy of $\sim$0.1 and $\sim$0.55 keV, respectively \cite[][]{kwkk07,pav05}, below the observed value of $0.7$ keV, but a possibility is that the atmospheric composition comprises heavier elements for which atomic transitions (still poorly known) fall at slightly higher energies.

Thanks to the extensive and dense monitoring performed by \nic, we could obtain a phase connected timing solution extending over 4 years that reveals significant variations in the spin-down rate. These are most likely due to the presence of timing noise, as shown by the residuals in Figure~\ref{fig:phasecon}. Several indicators have been proposed to quantify the level of timing noise in pulsars (see, e.g. \citealt{nam19}). One of them is the quantity $\Delta (T_{obs})$, defined as 

\begin{equation}
   \Delta(T_{obs}) = {\rm log}\left( \frac{1}{6\nu}|\nuddot|T_{obs}^3 \right).
\end{equation}

\noindent
where $T_{obs}$ is the time span of the observations over which $\nuddot$ is measured  \citep{arz94}.
Another commonly used measure of the timing noise is given by 

\begin{equation}
 \sigma_{TN}^2 = \sigma_R^2-\sigma_W^2,
\end{equation}

\noindent
where $\sigma_R$ is the rms of the residuals of the quadratic time solution and $\sigma_W$ is the typical error of the pulse phases. 

Our timing solution results in   $\Delta_8\equiv\Delta(10^8\, \mathrm s)$ = $-0.601\pm0.007$  and $\sigma_{TN} = 6.33$ ms. 
These values can be compared with those obtained for radio pulsars and other classes of isolated neutron stars.  An empirical correlation between $\Delta_8$ and $\pdot$  predicts $\Delta_8 = -2.1$ \citep{arz94}, but it has a large scatter, consistent with the value we derived for Calvera. The expected value of  $\sigma_{TN}$, according to the scaling relation by \citet{sha10}, is 5.5 ms. Thus we can conclude that the timing noise of Calvera is higher than the average but still consistent with the distribution of values seen in normal radio pulsars with similar $\pdot$.

\citet{zan11} reported the presence of diffuse X-ray emission about 13$'$ west of Calvera, with spectral properties consistent with a supernova remnant and without counterparts at other wavelengths. For the distance of $d=3.3$ kpc implied by our best fit to the X-ray spectrum, the dimensions of this diffuse emission,  $\sim$15$\times$8 pc,  are reasonable for a SNR possibly associated to Calvera. However, the   measured proper motion indicates that the pulsar is moving toward the putative remnant, rather than away from it. Therefore, we believe there is no physical connection  between  these two objects. 

\section{Conclusions} \label{sect:conclusions}

We have analyzed the recent observations of  Calvera obtained with \nic\ to explore the possibility that the observed thermal emission comes from the whole NS surface. We found that our hydrogen atmosphere model, computed for a temperature distribution given by a dipolar magnetic field, is able to well reproduce both the spectra and the pulse profiles, provided that an additional harder component, likely resulting from hot spots at the magnetic poles, is included. We also confirmed the presence of  spectral features below 1 keV, that can be fitted, with the minimal number of parameters, as an absorption edge. 

Our results demonstrate that the X-ray pulsations can be well reproduced even in the presence of thermal emission from the whole surface. The observed flux implies a distance larger than 3 kpc, which makes Calvera still underluminous in  $\gamma$-rays ($<7\times10^{32}\times d_{\rm 3~kpc}^2$ erg s$^{-1}$, \citealt{hal13}), but  less so than previously suggested. 

The most  striking property of Calvera is its height above the Galactic disk. The  thermal luminosity of $1.4\times10^{33}$ erg s$^{-1}$  (for $d=3.3$ kpc) indicates that $\tau_c = P/2 \pdot\sim300$ kyr can not be too different from the true age. This supports the idea that  Calvera was born in the Galactic halo, most likely from the explosion of a run-away massive star or, possibly, in a more unusual event involving a halo star, such as, e.g., the accretion induced collapse of a white dwarf.

\acknowledgments 
We acknowledge  support via ASI/INAF Agreement n. 2019-35-HH and PRIN-MIUR 2017 UnIAM (Unifying Isolated and Accreting Magnetars, PI S.~Mereghetti).\\

\section*{Appendix}
The log of \nic\ observations used for the timing analysis are summarized in Table~\ref{tab:logtime}. The table lists the ObsIds, the start time in UTC units, the net exposure time after excluding periods of high particle background as described in Section~\ref{sec:data analysis}, and the number of the time interval in which each ObsId was grouped for the phase-connected timing analysis.

\startlongtable
\begin{deluxetable}{lccc}
\tabletypesize{\footnotesize}
\label{tab:logtime}
\tablecaption{Log of \nic\ observations}
\tablecolumns{4}
\tablewidth{0pc}
\tablehead{
\colhead{Observation}   &
\colhead{Start time} &
\colhead{Exposure} & 
\colhead{Grouping}\\
\colhead{Id}   &
\colhead{(UTC)} &
\colhead{(ks)} & 
\colhead{Number}
}
\startdata
1020290102   &   2017-09-18T05:03:19   &    1.9   &     1 \\
1020290103   &   2017-09-19T07:16:03   &    0.9   &     1 \\
1020290104   &   2017-09-20T03:19:27   &    1.4   &     1 \\
1020290105   &   2017-09-21T09:08:56   &    2.1   &     1 \\
1020290106   &   2017-09-22T04:43:09   &    5.8   &     2 \\
1020290107   &   2017-10-07T07:31:34   &   10.1   &     3 \\
1020290108   &   2017-10-08T02:05:01   &    9.3   &     4 \\
1020290109   &   2017-10-08T23:37:31   &   16.9   &     5 \\
1020290110   &   2017-10-10T00:19:35   &    6.5   &     6 \\
1020290120   &   2017-12-22T02:57:46   &    8.6   &     7 \\
1020290122   &   2018-02-21T00:27:28   &    7.0   &     8 \\
1020290123   &   2018-02-22T07:21:47   &    2.3   &     9 \\
1020290124   &   2018-02-23T09:32:53   &    1.0   &     9 \\
1020290125   &   2018-02-24T16:29:13   &    1.9   &     9 \\
1020290126   &   2018-02-25T03:19:24   &    1.3   &    10 \\
1020290127   &   2018-02-26T00:54:14   &    2.5   &    10 \\
1020290128   &   2018-02-27T05:56:32   &    4.8   &    10 \\
1020290129   &   2018-03-03T01:17:31   &   11.7   &    11 \\
1020290130   &   2018-03-04T03:15:26   &    8.4   &    12 \\
1020290131   &   2018-03-05T01:12:57   &    8.1   &    13 \\
1020290135   &   2018-03-26T10:48:44   &    2.0   &    14 \\
1020290136   &   2018-03-27T03:54:05   &    4.6   &    14 \\
1020290143   &   2018-04-06T01:32:29   &    7.8   &    15 \\
1020290144   &   2018-04-07T00:55:57   &    6.2   &    16 \\
1020290150   &   2018-04-27T00:35:18   &    5.4   &    17 \\
1020290152   &   2018-05-01T00:21:16   &   12.7   &    18 \\
1020290153   &   2018-05-13T05:26:08   &    9.6   &    19 \\
1020290154   &   2018-05-22T22:19:06   &    0.5   &    20 \\
1020290155   &   2018-05-23T01:24:31   &    1.6   &    20 \\
1020290156   &   2018-05-24T02:06:08   &    2.3   &    20 \\
1020290157   &   2018-05-26T03:29:27   &    3.3   &    20 \\
1020290158   &   2018-05-27T01:05:59   &    6.8   &    21 \\
1020290159   &   2018-05-28T00:15:12   &    6.4   &    22 \\
1020290160   &   2018-05-29T00:57:10   &    1.8   &    23 \\
1020290161   &   2018-06-01T01:30:10   &    1.1   &    23 \\
1020290162   &   2018-06-02T02:14:52   &    0.8   &    23 \\
1020290163   &   2018-06-03T13:48:10   &    0.6   &    23 \\
1020290164   &   2018-06-04T02:09:23   &    0.3   &    23 \\
1020290165   &   2018-06-05T02:51:18   &    0.7   &    23 \\
1020290166   &   2018-06-06T20:24:39   &    0.7   &    24 \\
1020290167   &   2018-06-07T05:54:44   &    0.4   &    24 \\
1020290168   &   2018-06-08T01:58:41   &    4.0   &    24 \\
1020290169   &   2018-06-09T02:38:47   &    1.6   &    25 \\
1020290171   &   2018-06-13T06:55:41   &    2.3   &    25 \\
1020290172   &   2018-06-13T23:55:00   &    2.3   &    25 \\
1020290174   &   2018-06-16T04:25:15   &    5.2   &    26 \\
1020290176   &   2018-07-02T03:09:13   &    3.2   &    27 \\
1020290177   &   2018-07-03T05:21:57   &    3.8   &    27 \\
1020290183   &   2018-07-23T15:51:42   &    0.4   &    28 \\
1020290184   &   2018-07-24T07:17:36   &    0.7   &    28 \\
1020290185   &   2018-07-26T17:58:10   &    2.2   &    28 \\
1020290186   &   2018-07-27T01:41:10   &    1.9   &    28 \\
1020290188   &   2018-07-28T23:58:04   &    5.0   &    29 \\
1020290189   &   2018-08-01T11:19:26   &    2.0   &    30 \\
1020290190   &   2018-08-03T15:52:37   &    0.6   &    30 \\
1020290191   &   2018-08-05T09:58:25   &    0.9   &    30 \\
1020290192   &   2018-08-06T07:36:04   &    2.0   &    30 \\
1020290194   &   2018-08-08T00:53:20   &    7.9   &    31 \\
1020290198   &   2018-08-12T00:38:12   &    5.1   &    32 \\
1020290199   &   2018-08-13T01:23:42   &    7.2   &    33 \\
1020290204   &   2018-08-17T00:57:45   &    7.9   &    34 \\
1020290206   &   2018-08-19T00:59:48   &    6.1   &    35 \\
1020290207   &   2018-08-23T03:43:40   &    0.2   &    36 \\
1020290208   &   2018-08-24T00:03:35   &    2.1   &    36 \\
1020290209   &   2018-08-25T00:32:41   &    4.8   &    36 \\
1020290211   &   2018-08-29T07:52:20   &    3.6   &    37 \\
1020290212   &   2018-09-03T08:32:14   &    1.0   &    37 \\
1020290213   &   2018-09-04T20:02:30   &    1.6   &    37 \\
1020290215   &   2018-10-03T01:18:39   &    6.7   &    38 \\
1020290216   &   2018-10-04T11:17:10   &    3.9   &    39 \\
1020290217   &   2018-10-06T11:01:49   &    1.8   &    39 \\
1020290218   &   2018-10-07T00:55:47   &    3.2   &    40 \\
1020290219   &   2018-10-08T04:46:10   &    0.6   &    40 \\
1020290220   &   2018-10-10T21:33:13   &    0.7   &    40 \\
1020290221   &   2018-10-11T00:38:17   &    1.3   &    40 \\
1020290225   &   2018-10-27T05:46:26   &    1.1   &    41 \\
1020290226   &   2018-10-28T07:57:42   &    2.7   &    41 \\
1020290227   &   2018-10-29T05:35:47   &    3.3   &    41 \\
1020290228   &   2018-10-30T03:13:47   &    3.1   &    42 \\
1020290229   &   2018-11-02T05:31:00   &    0.5   &    42 \\
1020290230   &   2018-11-03T06:09:54   &    2.2   &    42 \\
1020290233   &   2018-11-07T10:39:54   &    5.1   &    43 \\
1020290234   &   2018-11-08T11:25:29   &    1.5   &    44 \\
1020290235   &   2018-11-09T05:47:34   &    1.0   &    44 \\
1020290236   &   2018-11-11T14:57:21   &    2.5   &    44 \\
1020290238   &   2018-11-13T00:54:12   &    5.3   &    45 \\
1020290239   &   2018-11-14T12:23:47   &    0.4   &    46 \\
1020290240   &   2018-11-16T09:11:41   &    1.6   &    46 \\
1020290241   &   2018-11-17T06:48:58   &    1.4   &    46 \\
1020290242   &   2018-11-18T01:20:37   &    2.2   &    46 \\
1020290245   &   2018-12-06T04:10:03   &    5.7   &    47 \\
1020290247   &   2018-12-10T23:53:50   &    6.3   &    48 \\
1020290249   &   2018-12-15T08:47:21   &    7.2   &    49 \\
1020290256   &   2019-01-11T00:16:17   &    5.5   &    50 \\
1020290257   &   2019-01-12T08:45:57   &    1.2   &    51 \\
1020290258   &   2019-01-17T06:01:13   &    1.4   &    51 \\
1020290259   &   2019-01-18T02:02:14   &    4.6   &    51 \\
1020290263   &   2019-01-31T04:42:26   &    5.8   &    52 \\
1020290269   &   2019-02-10T08:41:18   &    5.2   &    53 \\
1020290270   &   2019-02-11T17:29:29   &    0.1   &    54 \\
1020290271   &   2019-02-12T02:28:43   &    4.6   &    54 \\
1020290272   &   2019-02-13T06:16:43   &    0.8   &    54 \\
1020290275   &   2019-02-19T01:11:20   &    7.4   &    55 \\
1020290276   &   2019-02-23T22:47:05   &    0.7   &    56 \\
1020290277   &   2019-02-24T00:19:45   &    3.3   &    56 \\
1020290278   &   2019-02-27T09:43:59   &    1.4   &    56 \\
1020290279   &   2019-02-28T05:53:00   &    1.4   &    57 \\
2020290201   &   2019-03-02T01:24:43   &    0.0   &    57 \\
2020290202   &   2019-03-06T07:08:14   &    2.4   &    57 \\
2020290203   &   2019-03-07T01:40:49   &    2.8   &    57 \\
2020290204   &   2019-03-08T11:42:22   &    5.3   &    58 \\
2020290205   &   2019-03-09T01:37:02   &    9.1   &    59 \\
2579020101   &   2019-04-02T14:05:31   &    9.0   &    60 \\
2020290210   &   2019-04-09T16:30:50   &    0.8   &    61 \\
2020290211   &   2019-04-13T09:35:39   &    4.2   &    61 \\
2579020201   &   2019-06-08T03:40:39   &    9.5   &    62 \\
2579020305   &   2019-08-24T23:56:10   &    7.1   &    63 \\
2579020306   &   2019-08-26T16:18:55   &    3.2   &    64 \\
2579020307   &   2019-08-26T23:57:11   &    1.1   &    64 \\
2579020308   &   2019-08-31T15:16:03   &    0.0   &    64 \\
2579020309   &   2019-09-01T19:19:10   &    1.7   &    64 \\
2579020401   &   2019-10-20T14:10:06   &    5.2   &    65 \\
2579020404   &   2019-11-01T04:56:23   &    5.7   &    66 \\
2579020405   &   2019-11-02T00:59:56   &    7.8   &    67 \\
2579020407   &   2019-11-04T00:59:43   &    7.7   &    68 \\
2579020409   &   2019-11-06T00:56:53   &    6.7   &    69 \\
2579020410   &   2019-11-07T00:15:10   &    1.8   &    70 \\
2579020411   &   2019-11-08T00:49:55   &    4.1   &    70 \\
2579020412   &   2019-11-09T09:26:55   &    0.5   &    71 \\
2579020413   &   2019-11-10T08:41:00   &    0.3   &    71 \\
2579020414   &   2019-11-12T00:52:51   &    0.8   &    71 \\
2579020415   &   2019-11-13T09:07:07   &    2.5   &    71 \\
2579020416   &   2019-11-14T02:22:27   &    0.8   &    71 \\
2579020417   &   2019-11-16T02:21:03   &    1.5   &    71 \\
2579020419   &   2019-11-25T16:44:14   &    0.2   &    72 \\
2579020420   &   2019-11-26T09:35:02   &    1.2   &    72 \\
2579020421   &   2019-11-27T01:03:22   &    0.2   &    72 \\
2579020422   &   2019-11-28T06:27:27   &    3.6   &    72 \\
2579020501   &   2019-12-07T08:32:35   &    1.1   &    73 \\
2579020502   &   2019-12-08T03:09:18   &    3.8   &    73 \\
2579020503   &   2019-12-09T05:37:08   &    4.8   &    73 \\
2579020505   &   2019-12-11T00:57:55   &    7.1   &    74 \\
2579020508   &   2019-12-15T00:59:22   &    7.5   &    75 \\
2579020512   &   2020-01-08T02:24:04   &    6.6   &    76 \\
2579020513   &   2020-01-09T00:04:39   &    3.3   &    77 \\
2579020514   &   2020-01-10T08:35:57   &    2.1   &    77 \\
2579020516   &   2020-01-12T02:11:11   &    9.1   &    78 \\
2579020517   &   2020-01-13T01:35:07   &    3.3   &    79 \\
2579020518   &   2020-01-14T09:57:16   &    1.0   &    79 \\
2579020519   &   2020-01-15T03:09:47   &    1.0   &    79 \\
2579020520   &   2020-01-24T14:21:06   &    0.5   &    80 \\
2579020521   &   2020-01-25T08:50:39   &    1.3   &    80 \\
2579020601   &   2020-01-27T04:12:48   &    1.6   &    80 \\
2579020602   &   2020-01-28T01:52:25   &    2.1   &    80 \\
2579020607   &   2020-02-18T00:55:08   &    6.4   &    81 \\
2579020609   &   2020-02-24T08:43:11   &    6.0   &    82 \\
3536030107   &   2020-04-16T08:40:12   &    5.4   &    83 \\
3536030114   &   2020-05-03T00:09:29   &    9.4   &    84 \\
3536030115   &   2020-05-04T00:57:33   &    2.4   &    85 \\
3536030116   &   2020-05-05T00:09:56   &    1.4   &    85 \\
3536030117   &   2020-05-07T18:37:27   &    1.2   &    85 \\
3536030118   &   2020-05-18T11:37:35   &    1.6   &    86 \\
3536030119   &   2020-05-19T06:15:19   &    0.3   &    86 \\
3536030120   &   2020-05-20T00:49:33   &    0.2   &    86 \\
3536030121   &   2020-05-22T00:48:00   &    1.0   &    86 \\
3536030122   &   2020-05-23T00:00:56   &    3.2   &    86 \\
3536030123   &   2020-05-26T03:57:54   &    1.8   &    87 \\
3536030124   &   2020-05-27T04:44:49   &    1.1   &    87 \\
3536030125   &   2020-05-28T02:26:13   &    2.2   &    87 \\
3536030126   &   2020-05-29T09:24:29   &    1.9   &    88 \\
3536030127   &   2020-05-30T02:20:30   &    3.6   &    88 \\
3536030129   &   2020-06-01T00:46:07   &   11.2   &    89 \\
3536030131   &   2020-06-03T00:50:29   &   10.0   &    90 \\
3536030132   &   2020-06-04T00:04:45   &    5.0   &    91 \\
3536030134   &   2020-06-06T04:43:17   &    7.6   &    92 \\
3536030135   &   2020-06-07T00:37:54   &    9.3   &    93 \\
3536030136   &   2020-06-08T01:31:43   &    6.0   &    94 \\
3536030201   &   2020-06-19T03:44:54   &    6.4   &    95 \\
3536030203   &   2020-06-21T13:14:59   &    5.0   &    96 \\
3536030204   &   2020-06-21T23:55:06   &    9.9   &    97 \\
3536030205   &   2020-06-23T02:17:35   &    7.7   &    98 \\
3536030206   &   2020-06-23T23:54:29   &    5.8   &    99 \\
3536030207   &   2020-06-25T03:51:26   &    2.1   &   100 \\
3536030208   &   2020-06-26T09:04:34   &    4.8   &   100 \\
3536030209   &   2020-06-27T09:48:52   &    4.6   &   101 \\
3536030210   &   2020-06-27T23:46:47   &    1.8   &   101 \\
3536030211   &   2020-06-29T03:39:25   &    6.4   &   102 \\
3536030212   &   2020-07-01T18:56:56   &    0.6   &   103 \\
3536030213   &   2020-07-02T21:46:37   &    0.0   &   103 \\
3536030214   &   2020-07-03T14:21:23   &    1.7   &   103 \\
3536030215   &   2020-07-04T01:16:51   &    3.5   &   103 \\
3536030218   &   2020-07-07T01:57:59   &   10.8   &   104 \\
3536030224   &   2020-07-26T22:42:02   &    0.3   &   105 \\
3536030225   &   2020-07-28T05:41:30   &    0.2   &   105 \\
3536030226   &   2020-07-29T20:25:51   &    0.5   &   105 \\
3536030227   &   2020-07-30T11:55:12   &    1.6   &   105 \\
3536030228   &   2020-07-31T09:36:11   &    3.5   &   105 \\
3536030229   &   2020-08-01T10:19:31   &    1.5   &   106 \\
3536030230   &   2020-08-02T09:34:25   &    0.8   &   106 \\
3536030231   &   2020-08-03T10:22:11   &    0.3   &   106 \\
3536030232   &   2020-08-07T07:13:42   &    2.4   &   106 \\
3536030234   &   2020-08-18T14:38:08   &    1.4   &   107 \\
3536030235   &   2020-08-19T00:59:34   &    3.7   &   107 \\
3536030236   &   2020-08-20T03:42:17   &    0.4   &   108 \\
3536030237   &   2020-08-20T23:38:54   &    1.5   &   108 \\
3536030238   &   2020-08-22T02:16:38   &    0.9   &   108 \\
3536030239   &   2020-08-24T01:50:23   &    3.6   &   108 \\
3536030309   &   2020-09-22T22:38:06   &    0.3   &   109 \\
3536030310   &   2020-09-23T03:16:29   &    2.3   &   109 \\
3536030311   &   2020-09-24T21:01:30   &    0.6   &   109 \\
3536030312   &   2020-09-25T01:39:45   &    3.1   &   109 \\
3536030313   &   2020-09-26T00:54:51   &    5.0   &   110 \\
3536030314   &   2020-09-27T14:06:46   &    0.5   &   111 \\
3536030315   &   2020-09-28T00:55:42   &    4.0   &   111 \\
3536030316   &   2020-09-29T06:22:00   &    1.0   &   111 \\
3536030317   &   2020-10-02T19:31:06   &    0.3   &   112 \\
3536030318   &   2020-10-03T10:57:26   &    4.5   &   112 \\
3536030319   &   2020-10-04T17:57:14   &    2.5   &   112 \\
3536030320   &   2020-10-05T00:08:51   &    3.2   &   113 \\
3536030321   &   2020-10-10T10:15:53   &    1.8   &   113 \\
3536030401   &   2020-10-16T16:22:35   &    6.1   &   114 \\
3536030402   &   2020-10-24T21:14:12   &    0.7   &   115 \\
3536030403   &   2020-10-25T01:59:55   &    0.6   &   115 \\
3536030404   &   2020-10-28T21:17:54   &    0.7   &   115 \\
3536030405   &   2020-10-29T03:31:59   &    4.9   &   115 \\
3536030406   &   2020-10-30T02:30:30   &    8.4   &   116 \\
3536030407   &   2020-11-01T09:04:56   &    5.4   &   117 \\
3536030408   &   2020-11-02T00:24:41   &    5.3   &   118 \\
3536030409   &   2020-11-04T01:55:11   &   11.1   &   119 \\
3536030411   &   2020-11-06T05:01:56   &    7.3   &   120 \\
3536030413   &   2020-11-09T00:45:36   &    5.6   &   121 \\
3536030502   &   2020-12-09T00:42:47   &   11.0   &   122 \\
3536030507   &   2021-01-01T01:37:17   &    9.3   &   123 \\
3536030509   &   2021-01-08T02:13:42   &   14.1   &   124 \\
3536030602   &   2021-02-18T09:14:55   &    9.7   &   125 \\
3536030604   &   2021-02-23T08:13:01   &    5.7   &   126 \\
\enddata
\end{deluxetable}




\bibliographystyle{aasjournal} 

%





\end{document}